# Spatio-temporal estimation of wind speed and wind power using extreme learning machines: predictions, uncertainty and technical potential

Federico Amato[1] · Fabian Guignard[2] · Alina Walch[3] · Nahid Mohajeri[4] · Jean-Louis Scartezzini[3] · Mikhail Kanevski[5]

**Abstract**
With wind power providing an increasing amount of electricity worldwide, the quantification of its spatio-temporal variations and the related uncertainty is crucial for energy planners and policy-makers. Here, we propose a methodological framework which (1) uses machine learning to reconstruct a spatio-temporal field of wind speed on a regular grid from spatially irregularly distributed measurements and (2) transforms the wind speed to wind power estimates. Estimates of both model and prediction uncertainties, and of their propagation after transforming wind speed to power, are provided without any assumptions on data distributions. The methodology is applied to study hourly wind power potential on a grid of $250 \times 250$ m$^2$ for turbines of 100 m hub height in Switzerland, generating the first dataset of its type for the country. We show that the average annual power generation per turbine is 4.4 GWh. Results suggest that around 12,000 wind turbines could be installed on all 19,617 km$^2$ of available area in Switzerland resulting in a maximum technical wind potential of 53 TWh. To achieve the Swiss expansion goals of wind power for 2050, around 1000 turbines would be sufficient, corresponding to only 8% of the maximum estimated potential.

**PAC Codes** 05.45.Tp · 89.60.-k · 89.20.Kk

## 1 Introduction

Climate change is universally recognized as one of the major challenges humanity will have to face over the next decades. Thus, the development of renewable energy systems plays a crucial role in many strategic frameworks for sustainable development (Rogelj et al. 2015; Amato et al. 2020a). This includes not only the Sustainable Development Goals (SDGs) defined by the United Nations, but also the ensemble of renewable energy targets defined by different jurisdictions, such as USA (Barbose et al. 2016), Europe (Oberthür 2010; Santopietro and Scorza 2021), India (Bhushan and Gopalakrishnan 2021), Switzerland (Prognos 2012). The importance of decarbonizing energy systems is easily understandable (McCollum et al. 2018). However, the transformation of energy systems poses technical and logistical challenges, which may imply major threats for many societal and environmental aspects (Kiesecker et al. 2019). Power plants, including wind turbines, often require large amounts of land, hence generating conflicts with other priority targets of sustainable development, such as the limitation of land take (Saganeiti et al. 2020), the increase in local agricultural productivity (Martellozzo et al. 2018), and the protection of biodiversity (Yenneti et al. 2016). The presence of such conflicts may be underestimated or overshadowed by the urgency of operating on energy networks to reduce their cost in terms

Federico Amato and Fabian Guignard have contributed equally to this work.

Extended author information available on the last page of the article of carbon dioxide production (Spillias et al. 2020). For these reasons, a proper planning of the expansion of renewable energy technologies is required to optimize the future location of power plants by considering precise estimates of power generation while taking into account the conflicts between the installation of power plants, nature and environmental protection. Among renewable resources wind energy is a promising one, potentially contributing to the energy transition in many parts of the world. In contrast to solar energy, it is available at any time of the day; however, it is highly variable and complex to model. Thus, the quantification of the spatial and temporal variation of wind power and the related uncertainty may provide valuable information for energy planners and policymakers. To date, most of the estimates in the domain of wind power generation are based on averaged annual wind speed models, which however can only be used as an indicator of the power generation potential in a geographic area. Indeed, it has been shown that the use of average annual wind speed models underestimated wind power generation (Nelson and Starcher 2018). Furthermore, the estimation of wind speed at (sub-)hourly frequency is essential to assess complementarities with other renewable resources, such as solar energy, or potential storage requirements for energy systems with high shares of wind power (Kruyt et al. 2017). Therefore, a methodology to precisely estimate the wind speed based on hourly time steps is needed.

Wind speed measurements are generally collected by sparsely located meteorological stations, and hence do not provide the uniform spatial coverage to estimate the power generation potential over large geographical regions at high spatial resolution. However, a high spatial resolution is necessary for accurate renewable resource assessments and for the evaluation of potential locations of future wind farms. Several methods have been developed to obtain wind speed values at locations where no measurements are available (Landberg et al. 2003). These can be broadly classified into physical—or deterministic—and statistical approaches. Physical models, such as the non-hydrostatic weather prediction or the Reynolds-averaged Navier–Strokes ones, are mostly based on the study of wind via the use of fluid dynamics equations. While this family of models can ensure good estimates, it generally has limitations in the use of large amount of data and in its large computational burdens. These limitations are particularly inconvenient when working with data collected over long time periods and in relatively large geographical areas.

Statistical approaches are used to model wind speed using its statistical relationship with a set of geo-environmental and topographical predictors. They include a wide range of models, from classical geostatistics to machine learning (ML) (Mosavi et al. 2019). The latter have become extremely popular over the last decades, as they can deal with the non-linearity of wind speed and take advantage of big data (Deng et al. 2021; Sasser et al. 2021). Another potential advantage of statistical methods is that they may enable the estimation of the uncertainty of wind speed prediction, which is very important for exploring the potential location of new wind farms (Mahmoud et al. 2018). Indeed, the uncertainty of wind assessment is a major factor influencing the investment risk related to the installation of wind power plants (Veronesi et al. 2015). This uncertainty has been sometimes estimated by imposing a distributional shape to wind speed measurements a priori (Veronesi et al. 2016; Laib et al. 2018). However, it could be more convenient to determine a procedure to estimate uncertainty without making any assumption on the distributional properties of wind data. Moreover, when the aim is to estimate wind power, the propagation of such uncertainty in the process of transformation of wind speed into power must also be considered. ML has been successfully applied to model wind speed at several spatial scales in different parts of the globe (Lai et al. 2020). Nonetheless, most applications focused on lower frequency than the hourly one, dealing with the modelling of daily or monthly means (Veronesi et al. 2017; Douak et al. 2013). Moreover, the approaches discussed in the literature always consider the spatial and temporal dimension of the wind speed patterns separately, hence producing models that account only for the spatial or the temporal correlation in data, respectively (Cellura et al. 2008; Xiao et al. 2018; Liu et al. 2020). To the best of our knowledge, no ML-based methodology has been proposed to solve spatio-temporal interpolation problems of wind speed while also accounting for prediction uncertainty and its propagation to the wind power estimation.

To address this gap, we propose a novel methodological framework to estimate time series of wind speed and wind power on a regular spatial grid. The framework consists of two main steps: First, the spatio-temporal field of wind speed is reconstructed from spatially irregularly distributed wind speed measurements. To this aim, we adapt the method previously proposed in Amato et al. (2020b) to include uncertainty estimation. The method decomposes the wind speed data into temporally referenced basis functions and their corresponding spatially distributed coefficients. By using an Extreme Learning Machine (ELM) ensemble algorithm to model the latter coefficients, the adapted method allows to estimate both model and prediction uncertainty without any assumption on data distributional patterns. The ELM-based uncertainty estimation, which was introduced in Guignard et al. (2021), is expanded in this work by considering the spatio-temporal nature of the data. Second, the spatio-temporal wind speed estimations are transformed to wind power using empirical models and the uncertainty is propagated through these models. Moreover, we formalise the propagation of uncertainty through the non-linear wind power models.

The methodology is applied to the study of wind power potential in Switzerland, where the complex orography makes wind modelling an extremely challenging task. Previous studies have attempted to model wind speed in this country, although focusing on monthly frequencies or without investigating prediction uncertainties and their propagation to the power generation potential (Robert et al. 2013; Assouline et al. 2019). Both these aspects are considered here, and data at higher frequency are used. For the application, 10 years of wind speed monitored data collected at an hourly frequency on a set of up to 208 monitoring stations have been investigated. Using the proposed two-step framework, the wind speed and its uncertainty are estimated for a grid of $250 \times 250$ m$^2$, and the wind power potential is derived for horizontal-axis wind turbines of 100 m hub height. The results are validated against past turbine generation data and compared to an existing wind speed estimation for Switzerland. We further quantify the national technical wind power potential, accounting for regulatory planning limitations related to noise abatement and natural, ecological and cultural heritage protection. This technical potential is assessed in the context of Switzerland's Energy Strategy, which aims at carbon neutrality by 2050 (BFE 2020). The strategy targets an annual wind power generation of 4.3 TWh to complement solar energy and replace existing nuclear power plants. While the analyses presented in the paper use annual values only, the 10-year hourly time series at high spatial resolution provided unprecedented opportunities for a wide range of spatio-temporal energy system assessments. By overlaying the results with spatial constraints and flexibly aggregating them at different spatial scales, the presented data can for example be integrated into the increasingly complex national energy system models aiming at optimizing the future Swiss electricity generation.

## 2 Methodology

This section presents the proposed framework to model wind speed and wind power generation potential, which consists of two steps: First, the wind speed data is interpolated from an irregularly-spaced monitoring network to a regular spatio-temporal field and the model and prediction uncertainties are estimated. Following Amato et al. (2020b), we show that a basis function representation can be used to consider the spatio-temporal dependencies in wind speed data by decomposing them into fixed temporal bases and stochastic spatial coefficients (Sect. 2.1). In contrast to previous work, the latter is modelled using an ensemble of Extreme Learning Machines (ELM) that permit uncertainty estimates. To estimate the model and prediction uncertainty, the method proposed by Guignard et al. (2021) is expanded to account for spatio-temporal nature of the data without making any assumption on data distributional patterns (Sect. 2.2). Second, we show how wind speed estimates and their corresponding uncertainties can be used to estimate the potential wind power generation (Sect. 2.3).

### 2.1 Spatio-temporal modelling of irregularly spaced data

This subsection describes the methodology to decompose spatio-temporal data via basis functions and to spatially model the resulting linear coefficients.

#### 2.1.1 Basis function decomposition of spatio-temporal data

Spatio-temporal wind speed observations collected by irregularly spaced monitoring stations, can be decomposed in a linear combination of purely temporal bases through principal component analysis (PCA), also known as empirical orthogonal function (EOF) analysis in the fields of meteorology and climatology (Cressie and Wikle 2011). The linear coefficients of the combination, which will be modelled, are purely spatial.

Assume that we have spatio-temporal measurements $\{Z(\mathbf{s}_i, t_j)\}$ at $S$ locations $\{\mathbf{s}_i : 1 \leq i \leq S\}$ and $T$ times $\{t_j : 1 \leq j \leq T\}$, with $S \leq T$. Let us define the *empirical temporal mean* at time $t_j$ by

$$\widehat{\mu}_t(t_j) := \frac{1}{S} \sum_{i=1}^{S} Z(\mathbf{s}_i, t_j), \tag{1}$$

and the *temporally centered data* by

$$\widetilde{Z}(\mathbf{s}_i, t_j) := Z(\mathbf{s}_i, t_j) - \widehat{\mu}_t(t_j). \tag{2}$$

Then, the temporally centered data can be written as

$$\widetilde{Z}(\mathbf{s}_i, t_j) = \sum_{k=1}^{S} a_k(\mathbf{s}_i) \phi_k(t_j), \tag{3}$$

where the $\phi_k(t_j)$ form a discrete orthonormal temporal basis and the $a_k(\mathbf{s}_i)$ are the spatial coefficients with respect to the $k$-th EOF $\phi_k$ at locations $\mathbf{s}_i$, such that

$$\begin{aligned}
&\mathbb{E}[a_k(\mathbf{s}_i)] = 0, \text{ for all } k \text{ and all } i, \\
&\text{Var}[a_k(\mathbf{s}_i)] \geq \text{Var}[a_{k+1}(\mathbf{s}_i)] \geq 0, \text{ for all } k \text{ and all } i, \\
&\text{Cov}[a_k(\mathbf{s}_i), a_l(\mathbf{s}_i)] = 0, \text{ for all } k \neq l \text{ and all } i.
\end{aligned} \tag{4}$$

The spatio-temporal measurements are then supposed to follow

$$Z(\mathbf{s}_i, t_j) = \widehat{\mu}_t(t_j) + \sum_{k=1}^{S} a_k(\mathbf{s}_i)\phi_k(t_j) + \eta(\mathbf{s}_i, t_j), \quad (5)$$

where $\eta(\mathbf{s}_i, t_j)$ is an error term with zero mean, which includes any stochastic part which is not described by the model and may contain spatio-temporal dependencies.

The basis is obtained by a spectral decomposition of the empirical temporal covariance matrix, from which temporal EOFs with spatial coefficients are obtained. However, in this application, it was computed by a singular value decomposition, which is more beneficial from a computational perspective. Several practical considerations can be found in Wikle et al. (2019).

### 2.1.2 Extreme learning machine

ELM is a fast and efficient single-layer feedforward neural network (Huang et al. 2006). The input weights and biases are randomly chosen, and the output weights are optimized through least-squares. ELM can address spatial interpolation tasks and deal with high-dimensional environmental data (Leuenberger and Kanevski 2015).

Denoting the transpose operator as $(\cdot)^T$, suppose that $d$ input variables $\mathbf{x} = (x_1, \ldots, x_d)^T \in \mathbb{R}^d$ are related to an output variable $y \in \mathbb{R}$ through the relationship

$$y = f(\mathbf{x}) + \varepsilon(\mathbf{x}), \quad (6)$$

where $f(\mathbf{x})$ is a function and $\varepsilon(\mathbf{x})$ is a centred random noise with finite variance, both depending on the input.

Let $\{(\mathbf{x}_i, y_i) : \mathbf{x}_i \in \mathbb{R}^d, y_i \in \mathbb{R}\}_{i=1}^{n}$ be a training set. Given $N$, the number of neurons of the hidden layer, the input weights $\mathbf{w}_j \in \mathbb{R}^d$ and biases $b_j \in \mathbb{R}$ are randomly initialized for $j = 1, \ldots N$. In this paper, all input weights and biases are independently and uniformly drawn between $-1$ and 1. The $n \times N$ hidden layer matrix, denoted as $\mathbf{H}$, is defined element-wise by $\mathbf{H}_{ij} = g(\mathbf{x}_i^T \mathbf{w}_j + b_j)$, $i = 1, \ldots, n$, and $j = 1, \ldots, N$, where $g$ is an infinitely differentiable activation function. Here, the logistic function is chosen as an activation function.

The function $f(\mathbf{x})$ is supposed to be related to the hidden matrix by $f(\mathbf{x}) = \mathbf{H}\boldsymbol{\beta}$, where $\boldsymbol{\beta}$ is the vector of output weights. The output weights $\boldsymbol{\beta}$ are then estimated using least squares. A regularized version of ELM is used here (Deng et al. 2009), with the benefits of stabilizing the variability of the output weights and reducing overfitting and outliers effects. This corresponds to minimizing the cost function

$$J(\boldsymbol{\beta}) = \|\mathbf{y} - \mathbf{H}\boldsymbol{\beta}\|_2^2 + \alpha\|\boldsymbol{\beta}\|_2^2, \quad (7)$$

for some fixed $\alpha > 0$, where $\|\cdot\|_2$ denotes the Euclidean norm and $\mathbf{y} = (y_1, \ldots, y_n)^T$. The real number $\alpha$ is sometimes called the *Tikhonov factor* and controls the amount of regularization. Noting $\mathbf{I}$ the identity matrix and $\mathbf{H}^\alpha = (\mathbf{H}^T\mathbf{H} + \alpha\mathbf{I})^{-1}\mathbf{H}^T$, the solution of this minimization problem is given by $\widehat{\boldsymbol{\beta}} = \mathbf{H}^\alpha \mathbf{y}$. This model is a ridge regression (Piegorsch 2015) performed on the random feature space (Lendasse et al. 2013). Then, given a new input point $\mathbf{x}_0 \in \mathbb{R}^d$, the prediction is given by $\widehat{f}(\mathbf{x}_0) = \mathbf{h}^T \widehat{\boldsymbol{\beta}}$, where

$$\mathbf{h} = \left(g(\mathbf{x}_0^T \mathbf{w}_1 + b_1), \ldots, g(\mathbf{x}_0^T \mathbf{w}_N + b_N)\right)^T. \quad (8)$$

To enable variance estimation of the ELM modelling, the algorithm is retrained $M$ times and averaged (Guignard et al. 2021), resulting in a particular case of ELM ensembles (Lendasse et al. 2013; Liu and Wang 2010). Denoting the $m$-th prediction as $\widehat{f}_m(\mathbf{x}_0)$ for $m = 1, \ldots, M$, the final prediction is then

$$\widehat{f}(\mathbf{x}_0) = \frac{1}{M}\sum_{m=1}^{M}\widehat{f}_m(\mathbf{x}_0) = \frac{1}{M}\sum_{m=1}^{M}\mathbf{h}_m^T \mathbf{H}_m^\alpha \mathbf{y}, \quad (9)$$

where $\mathbf{h}_m$ and $\mathbf{H}_m^\alpha$ are the analogous quantities defined previously for the $m$-th model. Considering the input variables as deterministic, the use of several ELMs allows to develop distribution-free estimates of variance in homoskedastic (constant noise variance) and heteroskedastic (non-constant noise variance) settings. Several estimates are proposed in Guignard et al. (2021). In this paper, the heteroskedastic estimate $\widehat{\sigma}_{S2}^2$ will be used within the spatio-temporal model variance estimation in Sect. 2.2. Additionally, the bias-reduced homoskedastic model variance estimate $\widehat{\sigma}_{BR}^2$ and its related noise variance estimate $\widehat{\sigma}_\varepsilon^2$ will be used in the spatio-temporal prediction variance estimation procedure. Those variance estimates are also provided for regularised ELM and are computed using the UncELMe python package (see Guignard et al. 2021 for more details on their derivation and implementation).

### 2.1.3 Spatio-temporal modelling via spatial interpolation of the coefficients

As mentioned above, the data are assumed to follow equation (5) based on Amato et al. (2020b). The coefficients $a_k(\mathbf{s}_i) = a_k(\mathbf{s}_i, \mathbf{x}_i)$ depend only on space, potentially through additional spatial features $\mathbf{x}(\mathbf{s}_i)$. In the case of wind speed estimation, these features may include terrain characteristics such as altitude, slope or aspect. Using the single output strategy proposed in Amato et al. (2020b), the coefficient maps can be modelled with any ML algorithm, including ELM. For the $k$th map, this implicitly supposes the existence of a function $f_k$ such that

$$a_k(\mathbf{s}_i) = f_k(\mathbf{s}_i) + \varepsilon_k(\mathbf{s}_i), \tag{10}$$

where $\varepsilon_k(\mathbf{s}_i)$ is assumed to be a stochastic noise with zero mean and finite variance. The estimated function is denoted as $\hat{f}_k(\mathbf{s}_i) = \hat{a}_k(\mathbf{s}_i)$ and is used as a spatially interpolated coefficient map. The spatio-temporal prediction at a new point $\mathbf{s}_0$ is then given by

$$\widehat{Z}(\mathbf{s}_0, t_j) = \widehat{\mu}_t(t_j) + \sum_{k=1}^{S} \hat{a}_k(\mathbf{s}_0)\phi_k(t_j). \tag{11}$$

## 2.2 Uncertainty quantification

Using Eqs. (5) and (11), the prediction error is given by

$$\begin{aligned} Z(\mathbf{s}_0, t_j) - \widehat{Z}(\mathbf{s}_0, t_j) &= \sum_{k=1}^{K} [a_k(\mathbf{s}_0) - \hat{a}_k(\mathbf{s}_0)]\phi_k(t_j) + \eta(\mathbf{s}_0, t_j) \\ &= \underbrace{\sum_{k=1}^{K} [f_k(\mathbf{s}_0) - \hat{f}_k(\mathbf{s}_0)]\phi_k(t_j)}_{\text{modelling error}} \\ &\quad + \sum_{k=1}^{K} \varepsilon_k(\mathbf{s}_0)\phi_k(t_j) + \eta(\mathbf{s}_0, t_j). \end{aligned} \tag{12}$$

The first term on the right hand side is the modelling error between the linear combination of true regression functions $f_k(\mathbf{s}_0)$ and the spatio-temporal combination of spatial estimates $\hat{f}_k(\mathbf{s}_0)$. The variance of the modelling error, denoted as $\sigma_C^2(\mathbf{s}_0, t_j)$ and referred to as *spatio-temporal model variance*, quantifies the model accuracy. The spatio-temporal model variance will be used to construct model standard-error bands.

The prediction error will also be considered to evaluate accuracy of the estimate with respect to the observed output. As the prediction error distribution is unknown and no assumptions are made on the noise distribution, a reliable prediction interval estimation is not obvious. We prefer here to quantify the *spatio-temporal prediction variance*, given by the variance of the prediction error,

$$\sigma_P^2(\mathbf{s}_0, t_j) = \text{Var}\left[Z(\mathbf{s}_0, t_j) - \widehat{Z}(\mathbf{s}_0, t_j)\right]. \tag{13}$$

### 2.2.1 Spatio-temporal model variance estimation

Let us denote the vector of training outputs of the $k$-th map as $\mathbf{y}_k$, where the the $i$th vector component is given by $a_k(\mathbf{s}_i)$. In a similar manner, $\boldsymbol{\varepsilon}_k$ denotes the vector given by the noise at the training points. Assuming that $\text{Cov}[\boldsymbol{\varepsilon}_k, \boldsymbol{\varepsilon}_l] = 0$ ensures that no additional variability comes from the spatial model interactions. Indeed, knowing the training input, note that for a single ELM and for all $k \neq l$,

$$\begin{aligned} \text{Cov}[\hat{f}_k(\mathbf{s}_0), \hat{f}_l(\mathbf{s}_0)] &= \text{Cov}[\mathbf{h}_k^T \mathbf{H}_k^\alpha \mathbf{y}_k, \mathbf{h}_l^T \mathbf{H}_l^\alpha \mathbf{y}_l] \\ &= \text{Cov}[\mathbf{h}_k^T \mathbf{H}_k^\alpha \mathbb{E}[\mathbf{y}_k], \mathbf{h}_l^T \mathbf{H}_l^\alpha \mathbb{E}[\mathbf{y}_l]] \\ &\quad + \mathbb{E}[\mathbf{h}_k^T \mathbf{H}_k^\alpha \text{Cov}[\mathbf{y}_k, \mathbf{y}_l] \mathbf{H}_l^{\alpha T} \mathbf{h}_l] \\ &= \mathbb{E}[\mathbf{y}_k]^T \text{Cov}[\mathbf{h}_k^T \mathbf{H}_k^\alpha, \mathbf{h}_l^T \mathbf{H}_l^\alpha] \mathbb{E}[\mathbf{y}_l] \\ &\quad + \mathbb{E}[\mathbf{h}_k^T \mathbf{H}_k^\alpha \text{Cov}[\boldsymbol{\varepsilon}_k, \boldsymbol{\varepsilon}_l] \mathbf{h}_l] \\ &= 0, \end{aligned} \tag{14}$$

where the law of total covariance is used in the second equality. This result may be generalised to the ELM ensemble as

$$\text{Cov}\left[[f_k(\mathbf{s}_0) - \hat{f}_k(\mathbf{s}_0)]\phi_k(t_j), [f_l(\mathbf{s}_0) - \hat{f}_l(\mathbf{s}_0)]\phi_l(t_j)\right] = 0. \tag{15}$$

While it seems reasonable to suppose $\text{Cov}[\boldsymbol{\varepsilon}_k, \boldsymbol{\varepsilon}_l] = 0$, this should be validated e.g. by looking at the empirical cross-covariance function or the cross-variogram of the training residuals.

The spatio-temporal model variance is now straightforward to compute. Using Eq. (15), one obtains

$$\begin{aligned} \sigma_C^2(\mathbf{s}_0, t_j) &= \text{Var}\left[\sum_{k=1}^{K} [f_k(\mathbf{s}_0) - \hat{f}_k(\mathbf{s}_0)]\phi_k(t_j)\right] \\ &= \sum_{k=1}^{K} \text{Var}[f_k(\mathbf{s}_0)\phi_k(t_j) - \hat{f}_k(\mathbf{s}_0)\phi_k(t_j)] \\ &= \sum_{k=1}^{K} \text{Var}[\hat{f}_k(\mathbf{s}_0)]\phi_k^2(t_j). \end{aligned} \tag{16}$$

The spatio-temporal model variance is hence obtained directly by a sum of the spatial component model variances weighted by the corresponding squared basis function. Therefore, $\sigma_C^2(\mathbf{s}_0, t_j)$ can be estimated by using variance estimate of each ELM ensemble model,

$$\hat{\sigma}_C^2(\mathbf{s}_0, t_j) = \sum_{k=1}^{K} \hat{\sigma}_{S2,k}^2(\mathbf{s}_0)\phi_k^2(t_j), \tag{17}$$

where $\hat{\sigma}_{S2,k}^2(\mathbf{s}_0)$ is the heteroskedastic estimate $\hat{\sigma}_{S2}^2$ of the modelled regression function of the $k$th spatial coefficient map, at the input point $\mathbf{s}_0$. The choice of the estimate is motivated by the convenient trade-off between computational efficiency and estimation effectiveness of $\hat{\sigma}_{S2}^2$, see Guignard et al. (2021).

### 2.2.2 Spatio-temporal prediction variance estimation

The variance functions $\sigma_P^2(\mathbf{s}_0, t_j)$ are sometimes obtained by modelling them as a function of the input features using the squared residuals (Ruppert et al. 2003), as the expectation of the squared residuals approximately corresponds to the prediction variance (Carroll and Ruppert 1988). Using the squared residuals to perform a regression hence yields a plausible estimate of prediction variance (Hall and Carroll 1989).

The training squared residuals are given by

$$R^2(\mathbf{s}_i, t_j) = \left(Z(\mathbf{s}_i, t_j) - \widehat{Z}(\mathbf{s}_i, t_j)\right)^2, \quad (18)$$

here also denoted as $R^2$ for short. The latter is used to train a new model. This new model may result in negative estimates of $\sigma_P^2(\mathbf{s}_0, t_j)$. Hence, positiveness of the modelled variance function is here ensured through exponentiation, folllowing Ruppert et al. (2003) and Heskes (1997). The logarithm of the squared training residuals of the first model are then used as a new training set to model the random variable $L = L(\mathbf{s}_0, t_j) = \log(R^2(\mathbf{s}_0, t_j))$ with mean $\mu_L(\mathbf{s}_0, t_j)$ and variance $\sigma_L^2(\mathbf{s}_0, t_j)$. This second spatio-temporal model follows the same pipeline as the first model, including the EOF data decomposition and the ELM modelling on each of the resulting component with the high-dimensional input space composed by the spatially referenced features. Its predicted value is noted $\widehat{L}(\mathbf{s}_0, t_j)$.

A second order Taylor expansion around $\mu_L$ is needed to retrieve the expected squared residuals back from their log-transform, following the equation

$$\exp(L) \simeq \exp(\mu_L) + \exp(\mu_L)(L - \mu_L) \\ + \frac{1}{2}\exp(\mu_L)(L - \mu_L)^2. \quad (19)$$

Expansion of a random variable function in the neighborhood of the random variable mean is known as the *delta method* in statistics (Oehlert 1992; Ver Hoef 2012). Taking the expectation on both sides yields

$$\mathbb{E}[R^2] = \mathbb{E}[\exp(L)] \\ \simeq \exp(\mu_L) + \frac{1}{2}\exp(\mu_L)\mathbb{E}\left[(L - \mu_L)^2\right] \quad (20) \\ = \exp(\mu_L)\left(1 + \frac{1}{2}\sigma_L^2\right).$$

This motivates the following estimation of the spatio-temporal prediction variance,

$$\hat{\sigma}_P^2(\mathbf{s}_0, t_j) = \exp(\hat{\mu}_L)\left(1 + \frac{1}{2}\hat{\sigma}_L^2\right), \quad (21)$$

with the prediction of the second spatio-temporal model $\hat{\mu}_L = \widehat{L}(\mathbf{s}_0, t_j)$ and its prediction variance estimate

$$\hat{\sigma}_L^2 = \hat{\sigma}_L^2(\mathbf{s}_0, t_j) = \sum_{k=1}^{K}\left[\hat{\sigma}_{BR,k}^2(\mathbf{s}_0) + \hat{\sigma}_{\varepsilon,k}^2\right]\phi_k^2(t_j) \\ = \sum_{k=1}^{K}\hat{\sigma}_{BR,k}^2(\mathbf{s}_0)\phi_k^2(t_j) + \sum_{k=1}^{K}\hat{\sigma}_{\varepsilon,k}^2\phi_k^2(t_j), \quad (22)$$

where $\hat{\sigma}_{BR,k}^2(\mathbf{s}_0)$—respectively the noise estimate $\hat{\sigma}_{\varepsilon,k}^2$—is the bias-reduced homoskedastic estimate $\hat{\sigma}_{BR}^2$—respectively $\hat{\sigma}_\varepsilon^2$—of the $k$th modelled spatial coefficient map of the second spatio-temporal model. Although the noise of each component is not necessarily homoskedastic, $\hat{\sigma}_L^2(\mathbf{s}_0, t_j)$ is a good estimate of $\sigma_L^2(\mathbf{s}_0, t_j)$ and is better than limiting the estimation of $\sigma_P^2(\mathbf{s}_0, t_j)$ to a first order Taylor expansion.

### 2.3 Wind power estimation

Let us denote the expectation and variance of the wind speed $Z(\mathbf{s}_0, t_j)$ at a given location and time as $\mu_Z$ and $\sigma_Z^2$. The wind speed $Z(\mathbf{s}_0, t_j)$ has been measured at a height $h_1$. Assume that the wind speed $V(\mathbf{s}_0, t_j)$ at wind turbine height $h_2$ can be estimated by the so-called *log-law*,

$$V(\mathbf{s}_0, t_j) = Z(\mathbf{s}_0, t_j) \cdot \frac{\ln\frac{h_2}{h_0}}{\ln\frac{h_1}{h_0}}, \quad (23)$$

where $h_0 = h_0(\mathbf{s}_0)$ is the terrain roughness depending on the location (Whiteman 2000). The expectation $\mu_V$ and variance $\sigma_V^2$ of $V$ are then given by

$$\mu_V = \mathbb{E}\big[V(\mathbf{s}_0, t_j)\big] = \mu_Z \cdot \frac{\ln\frac{h_2}{h_0}}{\ln\frac{h_1}{h_0}}, \\ \sigma_V^2 = \mathrm{Var}\big[V(\mathbf{s}_0, t_j)\big] = \sigma_Z^2 \left(\frac{\ln\frac{h_2}{h_0}}{\ln\frac{h_1}{h_0}}\right)^2. \quad (24)$$

The wind speed at the wind turbine height is then converted to power. Logistic functions have proven to be highly precise in fitting power curves, on simulated and manufacturers data (Bokde et al. 2018; Villanueva and Feijóo 2016). Assume that the power curve $P(v)$ of the turbine is a three-parameter logistic function

$$P(v) = \phi_1 S(v) \quad \text{with} \quad S(v) = \frac{1}{1 + \exp\left(\frac{\phi_2 - v}{\phi_3}\right)}, \quad (25)$$

see Fig. 1 for an example. The first and second derivative of the power curve are (Minai and Williams 1993)

$$P'(v) = \frac{\phi_1}{\phi_3} S(v)(1 - S(v)) \\ P''(v) = \frac{\phi_1}{\phi_3^2} S(v)(1 - S(v))(1 - 2S(v)). \quad (26)$$

Due to the non-linearity of the power curve, the expectation and variance of $P(V)$ are again approximated using the delta method (Oehlert 1992; Ver Hoef 2012). The second order Taylor expansion of the power around $\mu_V$ is

$$P(V) \simeq P(\mu_V) + P'(\mu_V)(V - \mu_V) + \frac{1}{2}P''(\mu_V)(V - \mu_V)^2 \quad (27)$$

Taking the expectation on both side,

$$\mathbb{E}[P(V)] \simeq P(\mu_V) + \frac{1}{2}P''(\mu_V)\mathbb{E}\left[(V - \mu_V)^2\right]$$
$$= \phi_1 S(\mu_V)\left[1 + \frac{1}{2\phi_3^2}(1 - S(\mu_V))(1 - 2S(\mu_V))\sigma_V^2\right] \quad (28)$$

The variance of $P(V)$ is obtained by computing the variance of its first order Taylor expansion, as higher moments are not available, such that

$$\text{Var}[P(V)] \simeq (P'(\mu_V))^2 \sigma_V^2 = \frac{\phi_1^2}{\phi_3^2} S^2(\mu_V)(1 - S(\mu_V))^2 \sigma_V^2. \quad (29)$$

Given the parameters $\phi_1, \phi_2$ and $\phi_3$, the expected value and variance of the wind turbine power at each location $\mathbf{s}_0$ and each time $t_j$ are estimated by substituting $\mu_Z$ and $\sigma_Z^2$ by $\hat{Z}$ and $\hat{\sigma}_P^2$ in Eq. (24), and plug them into eqs. (28) and (29).

Equation (29) implies that the variance is completely transformed by the logistic function, see also Fig. 1. Thus, when wind speed is high with a sufficiently small amount of variance, the estimate remains confidently in the plateau region of the logistic function, characterised by the maximum wind power. Consequently, the power variance is small—in accordance with Eq. (29)—indicating a high confidence in having the maximum of energy production. Similarly, when wind speed is low with a relatively low variance, the power is close to zero with high certainty. By contrast, when the wind speed is in the transition phase of the logistic function, even with a very small variance, the power is susceptible to fluctuate between its minimum and maximum value. This leads to a high variance of the power estimate—characterised by a high derivative of $P(v)$ in Eq. (29).

## 3 Case study and data

This section introduces the case study for wind power estimation in Switzerland. First, we discuss the structures and properties of the wind data used in the remainder of the paper. Specifically, both the wind speed data and the spatially-referenced features used as input for the ML modelling will be presented. Then, the ELM model training based on the methodology proposed in Sect. 2.1.2 as well as the application of the wind power model for wind turbines of 100 m hub height is explained. Finally, we quantify the available area for installing wind turbines, which is required to obtain a national-scale estimate of the technical wind power potential for Switzerland.

### 3.1 Study area and data availability

Wind speed measurements have been obtained from the IDAWEB web portal of the Swiss Federal Office of Meteorology and Climatology (MeteoSwiss). The data are collected from 450 monitoring stations measuring wind speed at 10 m above the ground level with a 10 min frequency from 00:00 AM of the 1st January 2008 to 11:50 PM of the 31st December 2017. The number of available monitoring stations significantly changes over the sampling period, with relevant growth in 2013 and 2017. Therefore, data have been temporally divided into the following three sets, each having an homogeneous number of stations as indicated in Table 1:

- from 1st January 2008 00:00 am to 31st December 2012 11:50 pm, which will be referred to as *MSWind 08-12*,
- from 1st January 2013 00:00 am to 31st December 2016 11:50 pm, which will be referred to as *MSWind 13-16*,
- from 1st January 2017 00:00 am to 31st December 2017 11:50 pm, which will be referred to as *MSWind 17*.

For each dataset, the stations with more than 10% of missing or negative values have been removed, together with those having more than 10% of zero values. The

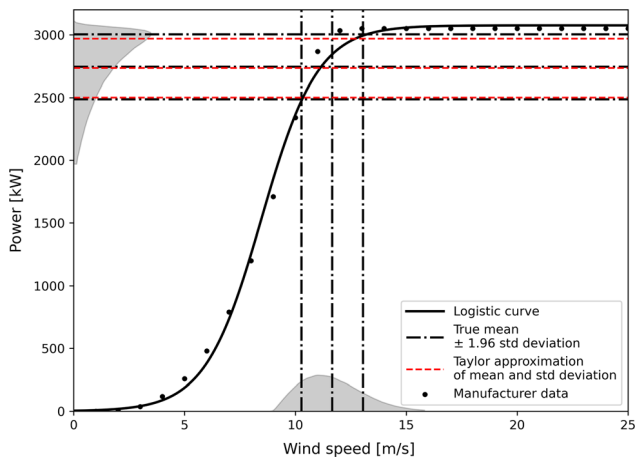

**Fig. 1** Logistic transformation and approximation of power generation with delta method. The power is considered as a logistic function of the wind speed, which is itself supposed to be a random variable. The distribution of wind speed is completely transformed by the non-linear function. Measurments for Enercon E-101 wind turbine from the manufacturer are also displayed

Table 1 Wind speed monitoring network datasets

|  | MSWind 08–12 | MSWind 13–16 | MSWind 17 |
| --- | --- | --- | --- |
| *General characteristics* | | | |
| Total number of stations | 106 | 127 | 208 |
| Number of training stations | 84 | 101 | 166 |
| Number of test stations | 22 | 26 | 42 |
| Time series length | 43′848 | 35′064 | 8′760 |
| *On the training datasets* | | | |
| Missing values | 2.1% | 1.1% | 1.2% |
| Minimal distance between stations (km) | ≤ 0.1 | 3.3 | 0.4 |
| Maximal distance between stations (km) | 323.1 | 332.7 | 332.7 |
| Characteristic network scale (km) | 22.2 | 20.2 | 15.8 |

Summary of the hourly wind speed datasets after preprocessing. When the monitoring network is not clustered, the characteristic network scale can be interpreted as an average distance between the monitoring stations and is computed as $\sqrt{A/n}$ where $A$ is the Switzerland area and $n$ is the number of stations in the network (Kanevski and Maignan 2004) remaining zero values have been set to missing values. Moreover, outliers and local suspicious behaviours, suggesting for example equipment failure, have been detected and replaced by missing values. The frequency of the data has then been reduced to 1 h by averaging.

Each of the three datasets has been divided into a training set (including 80% of the monitoring stations) and a test set (20%). Table 1 summarizes the main characteristics of the three cleaned datasets. Finally, all the remaining missing values of the training sets have been replaced by the local average data from the eight closer stations in space and the two contiguous time frames, yielding a mean over 24 spatio-temporal neighbours (Jun and Stein 2007; Porcu et al. 2016). Figure 2 indicates the location of the monitoring station in Switzerland, together with a division of the national territory into homogeneous geomorphological regions.

A full exploratory data analysis was performed on the three wind speed datasets and is available in "Appendix A.1 in Supplementary Material". The spatial plots in "Appendix A.1 in Supplementary Material" highlight the presence of structures related to the channelling effect and/or the climatic barrier formed by the alpine chain crossing the country. Time series plots and autocorrelation functions (ACF) have been used to identify the variety of temporal patterns in the data, including yearly and daily cycles with different intensities depending on the station. Finally, kernel density estimates (KDE) show how, while some stations seem to exhibit a Weibull distribution typical for wind speed measurements (Jung and Schindler 2019), many other stations are more atypical, sometimes even exhibiting bimodality. This highlights the importance of adopting a modelling approach which makes no distributional assumption on the data.

Wind speed has been proven to be extremely dependent on local orographic characteristics (Guignard et al. 2019), which can be assessed by applying convolutional filters to extract primary or secondary topographic features from a Digital Elevation Model (DEM) (Laib and Kanevski 2019). In this study, we adopted the 13-dimensional input space proposed in Robert et al. (2013) to model wind speed using ML. In addition to the coordinates of the geographical space (latitude, longitude and elevation), this input space includes three categories of spatial features:

- *Differences of Gaussians (DoG)* obtained by subtracting two smoothed surfaces attained through the application of Gaussian filters with different bandwidth to the DEM. Three different scales have been considered;
- *Directional derivatives* obtained evaluating the directional derivatives on DEMs smoothed with kernels having different bandwidth. Such filters are used to remove the spurious data of the DEMs, enhancing features in the data. Two scales have been considered for both North–South (N–S) and East–West (E–W) directions;
- *Terrain slopes* obtained as the norm of terrain gradient based on three smoothed DEMs.

Further details on the input features are provided in "Appendix A.2 in Supplementary Material".

### 3.2 Model training and application

#### 3.2.1 Wind speed

The modelling framework described in Sect. 2.1 has been applied to the MSWind 08-12, MSWind 13-16 and MSWind 17 datasets. For both the first and the second spatio-temporal model, the coefficients of each EOF

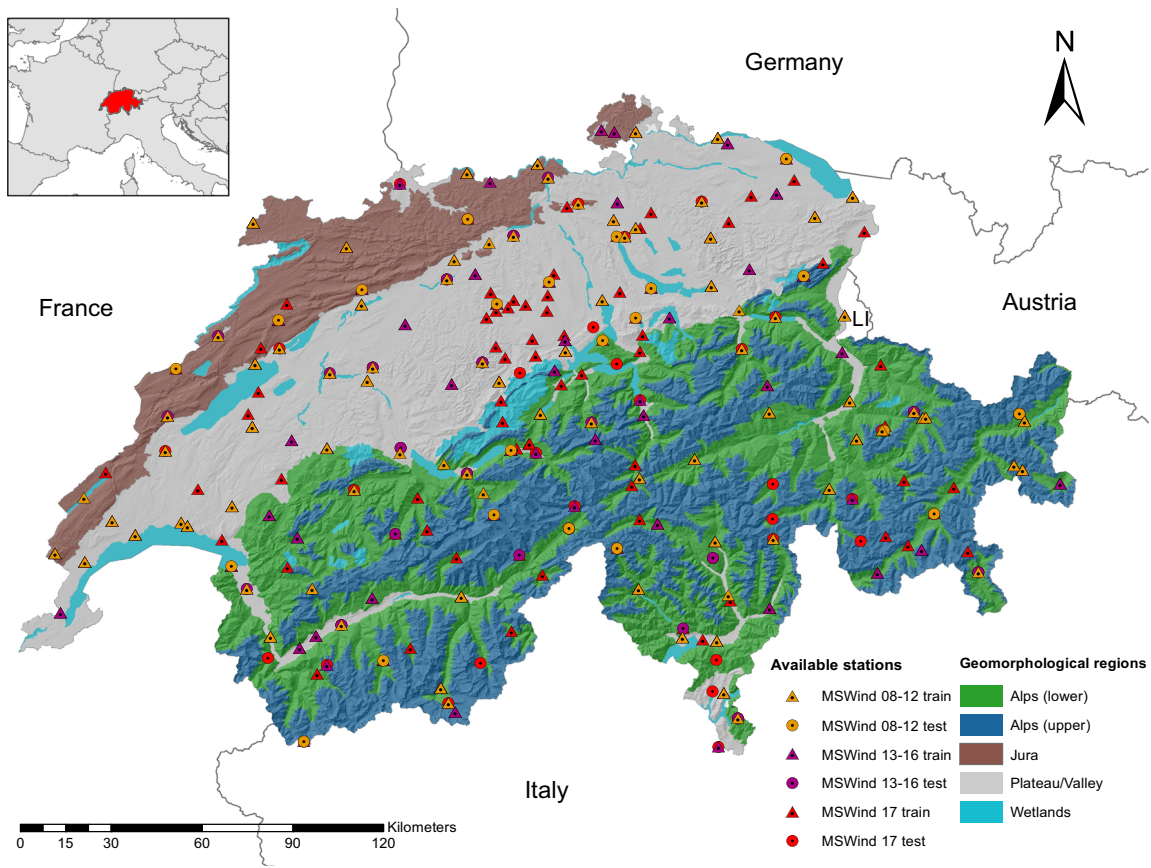

**Fig. 2** Study area and Swiss geomorphological regions. The location of the train and test stations belonging to MSWind 08–12, MSWind 13–16, and MSWind 17 is indicated within the different geomporhological units of the country

component have been spatially modelled with a regularised ELM ensemble of $M = 20$ members, with the 13-dimensional space presented in Sect. 3.1 as input features. Table 2 shows the number of neurons of each ELM ensemble—while it is fixed within each ensemble, it changes across the datasets to be slightly smaller than the number of training stations. This approach provides a high flexibility to the model. During model training, each member of each ELM ensemble is regularised by selecting a proper Tikhonov factor $\alpha$ via GCV (Golub et al. 1979; Piegorsch 2015). "Appendix B.1 in Supplementary Material" provides further details concerning model regularization. These include the use of the $\alpha$ values as indicator of the presence (or absence) of spatial structure in the modelled spatial coefficient maps, hence increasing the explainability of the ML model.

Test error metrics for the models are reported in Table 2, together with the time series of the empirical temporal means $\widehat{\mu}_t(t_j)$, computed from the training data and used as a prediction for the test stations. The latter are used as a baseline prediction benchmark. A comprehensive residuals analysis, here provided in the "Appendix B.2 in Supplementary Material", has been performed to verify the

**Table 2** Number of neurons, test RMSE and MAE

| Dataset | N | ST ELM model | | Emp. temp. mean | |
|---|---|---|---|---|---|
| | | RMSE | MAE | RMSE | MAE |
| MSWind 08–12 | 80 | 2.161 | 1.429 | 2.760 | 1.730 |
| MSWind 13–16 | 100 | 1.744 | 1.282 | 2.060 | 1.490 |
| MSWind 17 | 150 | 1.929 | 1.328 | 2.251 | 1.513 |

All the metrics are computed on the test set. The results based on the spatio-temporal ELM (ST ELM) model are benchmarked against the time series of the empirical temporal mean (Emp. temp. mean) $\widehat{\mu}_t(t_j)$ consistency of the obtained predictions and their uncertainty, highlighting the capability of the model to capture spatial, temporal and spatio-temporal dependencies in the data despite the complexity due to its hourly frequency and the relatively low number of training points.

Once trained, the models have been used to predict the spatio-temporal wind speed field and its model and prediction variances on a 250 m resolution regular grid, yielding three modelled spatio-temporal wind fields for Switzerland—one for each training dataset.

### 3.2.2 Wind power

To estimate the potential wind power generation in Switzerland, the approximated conversion and uncertainty propagation described in Sect. 2.3 are applied to the three modelled wind speed datasets. The power estimation is based on the characteristic parameters of an Enercon E-101 wind turbine at 100 m hub height (Enercon E-101 2021). The latter indicates the distance from the turbine platform to the rotor of an installed wind turbine, showing how high the turbine stands above the ground without considering the length of the turbine blades. Hence, the predicted wind speed data and its estimated variance are transformed from the measurement height of $h_1 = 10$ m to the hub height of $h_2 = 100$ m as described in Eq. (23), by considering a roughness $h_0$ derived from the Corine Land Cover (CLC)—issued from the Swiss Federal Office of Topography (SwissTopo)—following the methodology proposed in Grassi et al. (2015). Specifically, the CLC map of 2012 was used to estimate roughness for the MSWind 08–12 data, while the CLC map of 2018 was used for the two remaining datasets (further details are reported in "Appendix A.3 in Electronic suupplementary material"). In addition, all wind speeds greater than 25 m/s have been then discarded after the transformation. This value corresponds to the cut-out wind speed of the selected turbine as provided in the manufacturer's datasheet (Enercon E-101 2021). The manufacturer's wind turbine power curve (Enercon E-101 2021) has been fitted with the R Package WindCurves (Bokde et al. 2018), yielding $\phi_1 = 3075.31, \phi_2 = 8.47$ and $\phi_3 = 1.27$ for (25). Then, the transformed wind speed and its variance are passed into Eqs. (28) and (29). This yields an estimation of the expected electricity generation potential accompanied by its variance on the entire Switzerland over the 10 years from 2008 to 2017.

### 3.3 Available area for wind turbine installation

To convert the potential electricity generation per wind turbine into a national-scale potential estimate for wind power in the context of Switzerland's energy strategy, the available area for wind turbine installation and the potential number of turbines must be defined.

The available area for wind power installations is divided into four restriction zones, shown in Table 3, which indicate weather wind installation is (1) *prohibited*, (2) *restricted*, (3) inhibited by the presence of *forests*, or (4) no specific restrictions have been identified (*other*). These restriction zones are based on the framework for wind energy planning in Switzerland developed by the Swiss Federal Office of Spatial Development (ARE) (Bundesamt für Raumentwicklung 2020). Their exact definition is provided in "Appendix C.1 in Supplementary Material". In addition to the technical aspects considered here, the planning and installation of wind power plants is highly dependent on social, political and environmental concerns. We hence exclude only the *prohibited* zones for wind power installation. All other zones (*restricted, forests, other*) are used for the analysis in Sect. 4, whereby the different zones may be subject to different social, political or environmental considerations.

In the non-prohibited zones, wind turbines are virtually installed along the main direction of wind speed in Switzerland [SWW, 60° clockwise from north (Koller and Humar 2016)] using geospatial tools. To minimise the potential impact of one virtual turbine's generation on the next, turbines are spaced here by 16 turbine diameters (1.6 km) streamwise and 10 turbine diameters (1 km) spanwise. This is the double of the spacing that maximises the power output of a wind farm as assessed in Stevens et al. (2016), and agrees with the recommendations in Meyers and Meneveau (2012). The national-scale electricity potential is finally obtained as the electricity generation of each virtual turbine across the different restriction zones. While only annual values are considered in the analysis presented in this paper, hourly values may be used in future assessments of energy systems with high shares of wind power. Such assessments are beyond the scope of this work.

Table 3 Restriction zones for wind power in Switzerland

| Zone | Wind atlas | TLM regio | Altitude (m) |
| --- | --- | --- | --- |
| Prohibited | Building zones | Glaciers | > 3000 |
|  | Protected areas | Lakes |  |
|  | Areas excluded in principle |  |  |
| Restricted | Potential national interest | Protected areas | > 2500 |
| Forests | Inter-authority coordination, other | Forest |  |
| Other | Inter-authority coordination, other |  |  |

"Wind Atlas" denotes the restrictions as defined by ARE (Bundesamt für Raumentwicklung 2020), "swissTLM Regio" is the landscape model of Switzerland (Swisstopo 2020), and altitudes are derived from a digital terrain model (Swisstopo 2017)

# 4 Results

## 4.1 Wind speed modelling

Following the framework presented in Sect. 2, hourly predictions of wind speed were performed over the entire Swiss territory, covering the 10 years from 2008 to 2017. The top of Fig. 3 shows an example of predictions corresponding to January 2017 on a test station belonging to the MSWind 17 dataset. The model reproduces the main features of the measured wind speed time series, including most of the changes of magnitude and behaviour. However, the predicted time series appears smoother than the real data—this may be a consequence of the self-discarding of EOF components with a spatially unstructured coefficient map. Similar results are obtained for the MSWind 08–12 and the MSWind 13–16.

The estimation of the pointwise model and prediction standard-error bands, based respectively on $\pm 1.96\,\hat{\sigma}_C$ and $\pm 1.96\,\hat{\sigma}_P$, is also reported. The model standard-error band is quite narrow, suggesting a low variability of the mean prediction, despite the low number of training stations. By contrast, the prediction standard-error band is larger, as expected from the noisy nature of wind speed data. The true wind speed time series is hereby well encompassed in the $\pm 1.96$ prediction standard-error bands. For the same

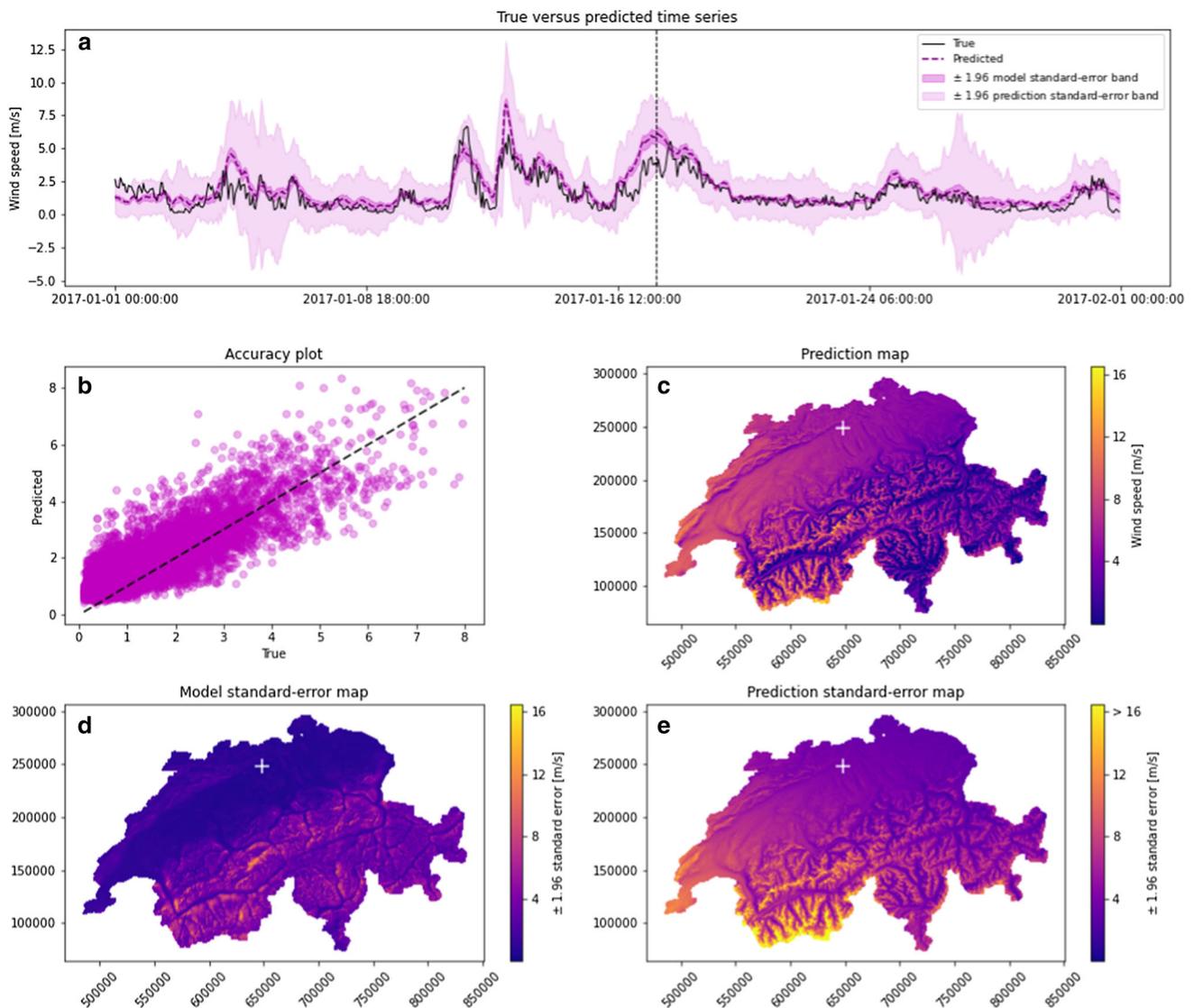

**Fig. 3** Model prediction of wind speed for the MSWind 17 dataset: **a** The true time series (in black) for a test station marked by a cross in the maps below and the predicted time series (in magenta). For visualisation purposes, only January 2017 is shown; **b** Accuracy plot at the same test station; **c** The predicted map of wind speed, at the fixed time indicated by the vertical dashed line in the temporal plot above; **d** Map of the model standard error multiplied by 1.96 at the same time; **e** Map of the prediction standard error multiplied by 1.96 at the same time

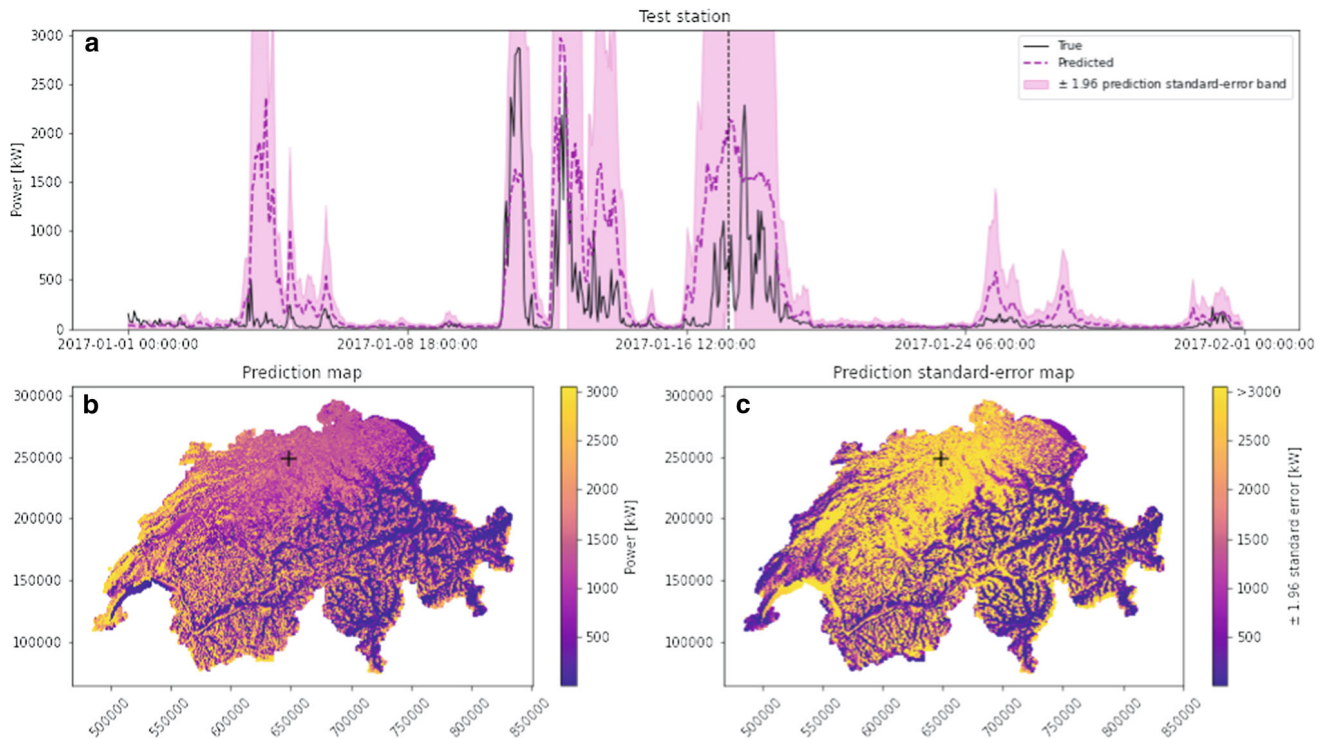

**Fig. 4** Model prediction for the MSWind 17 dataset: **a** The power time series at the same test station shown in Fig. 3 (in black) and the predicted time series (in magenta); **b** The predicted map of power generation, at the fixed time indicated by the vertical dashed line in the temporal plot above; **c** Map of the prediction standard error multiplied by 1.96 at the same time

test station, an accuracy plot is shown in the central row of the Fig. 3. Moreover, for the fixed time marked in the time series plot, a predicted map of wind speed is displayed. At the same fixed time, maps of the model and prediction standard-error are shown. Higher model and prediction variabilities are observed in the Alps, crossing the study region from the south-west to north-west. Qualitatively, it seems that the spatial scale of the pattern seen on the prediction standard-error map is comparable to the one observed on the prediction map, while the spatial scale of the pattern seen on the model standard-error map is coarser. This may be related to the multi-scale features used in the 13-dimensional input space.

### 4.2 Wind power estimation

The wind speed prediction at 10 m above ground have been used to estimate wind speed at 100 m; the latter estimates have then been transformed into wind power estimates following the methodology of Sect. 3.2.2. Figure 4 illustrates some samples of the results at the same location and period previously shown for wind speed modelling. The latter displays a partial power time series at a test station, a prediction map at a fixed time and its corresponding uncertainty quantification map. For comparison, the power obtained by passing the true wind speed measurement in the three-parameters logistic function $P(v)$ is added on the time series plots.

Generally speaking, the main behavioural variations of the true time series are captured and it is contained in the $\pm 1.96$ error bands. Interestingly, when production reaches its maximum potential defined by physical turbine characteristics, the error band sometimes shrinks. This was expected, due to the logistic transformation and its consequences on the variance behaviour stated previously. The maps provide a very interesting insight. An important part of the Jura region, in the north-western corner of the country, shows a very low uncertainty, while the power prediction is at its maximum. This behaviour is of particular interest for practical reasons, as it shows a high confidence of the model in these wind power estimates. Some similar spots are also identifiable in the western plateau.

The aggregation to annual total potential wind power generation, shown in Fig. 5 as the average value for the 10 years from 2008 to 2017, suggests that the potential is highest in the mountains, in both the Alps and the Jura, and may exceed 10 GWh in extreme cases. In the Plateau, the potential is lower (around 3–4 GWh), whereby zones with higher roughness length, such as urban areas, have a higher potential. Across the 10 years modelled in this work, the wind speed and wind power vary by up to 15–20% with respect to the 10-year mean (see Fig. 6). These variations may be explained through expected inter-annual variations of the meteorological conditions. Furthermore, differences in the number of weather stations used for the modelling may lead to variations in the estimated average wind speed. In particular, the large increase in training stations from 2016 to 2017 (101–166 stations) is expected to lead to a better representation of local weather patterns. Comparing the wind speed (left axis in Fig. 6) to the wind power (right axis) shows the impact of applying the logistic wind power curve, which increases the inter-annual variation of wind power.

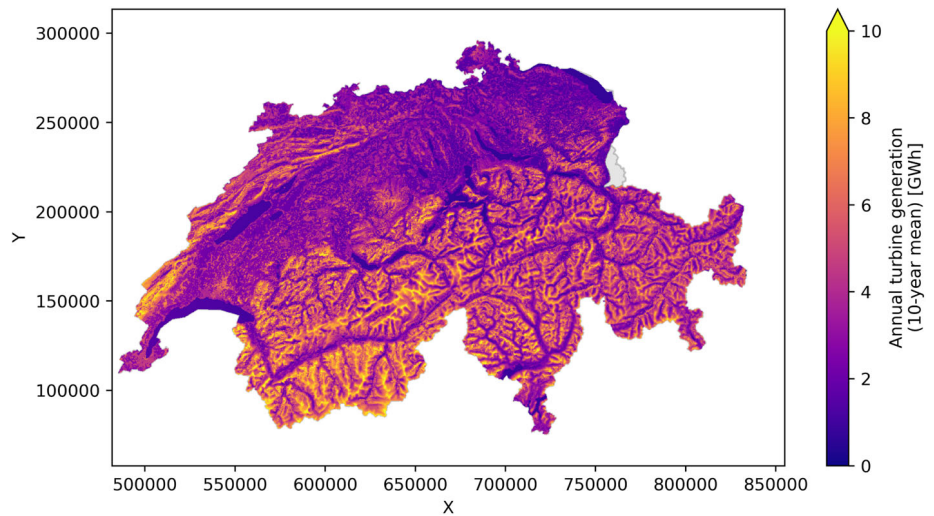

**Fig. 5** Annual potential wind power generation, averaged across all 10 years of measurement data

### 4.3 National-scale wind power potential in Switzerland

The application of the national-scale assessment of the available area for wind turbine installation (Sec. 3.3) shows that less than half of the surface of Switzerland may be considered for wind installations, as 52% of the area is in the *prohibited* zone. No particular restrictions have been identified for half of the remaining area, while the other half is either restricted or covered by forests (see Table 4). Assuming the occupation of 1.6 km$^2$ by each wind turbine, around 12,000 turbines could be installed if all available area (*restricted + forests + other*) was exploited. As Fig. 7 shows, much of the prohibited area is located in the Swiss Plateau (see Fig. 2 for reference), due to the high building density in this part of the country. The eastern Alps and the Jura mountains, on the other hand, show a lot of available area.

The average potential of these restriction zones for each part of Switzerland (Fig. 8a) shows that the mountain areas (upper and lower Alps, Jura) have the highest average wind power potential. The lowest potential is found in the Plateau and in mountain valleys, confirming the observations from Fig. 5. The annual average potential however only is one relevant aspect in the assessment of wind potential. Other factors related to the hourly wind power time series, such as generation peaks, the number of full-load hours or the average intra-day and seasonal variation of the potential, may also be derived from the results and represent relevant subjects of further work.

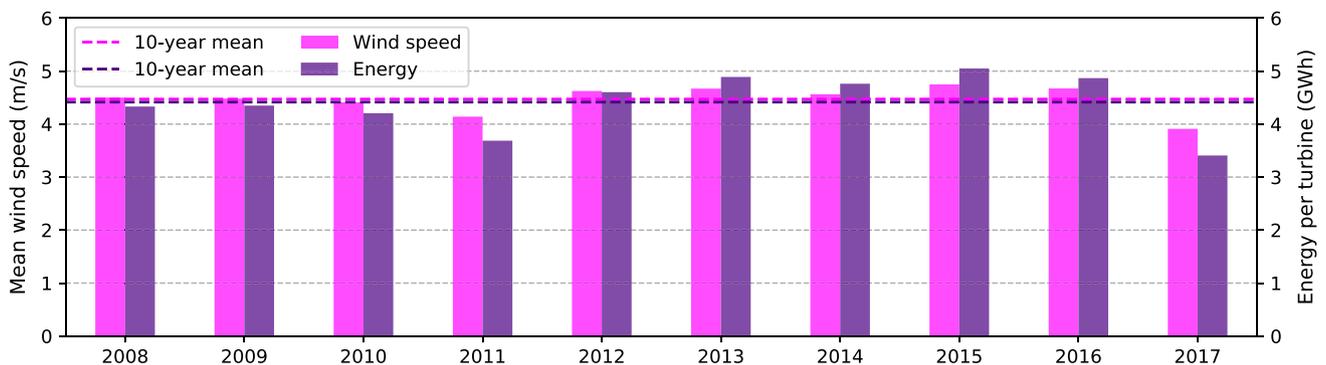

**Fig. 6** Annual mean wind potential: wind speed (left axis) and potential wind power generation (right axis) per potential turbine for the 10 years from 2008 to 2017

Table 4 Available area for wind turbine installation. Area covered, number of virtual turbines installed and cumulative annual wind potential for each restriction zone

| Zone | Area km$^2$ | Virtual turbines | Wind potential (TWh) |
| --- | --- | --- | --- |
| Prohibited | 21,672 (52%) | – | 0 |
| Restricted | 4351 (11%) | 2734 | 13.6 |
| Forests | 5315 (13%) | 3311 | 15.8 |
| Other | 9953 (24%) | 5985 | 23.7 |

Across the different restriction zones, the *other* zones have the lowest potential per turbine in all parts of Switzerland, followed by the *restricted* areas and *forests*. This may be explained by the fact that *other* zones are located at lower altitudes and in more flat terrain, yielding lower potentials. The high estimates potential of *forests* may be related to the higher roughness length in these zones. In the upper Alps, *restricted* areas have the highest average potential, likely due to their locations at higher altitudes with higher wind speeds.

Summing the potential across all virtual wind turbines (Fig. 8b) shows that the Alps make up for around 70% (36 TWh) of the national total potential of around 53 TWh. Half of this potential is located in the *other* zones and may hence be exploitable without specific geographic restrictions. In the lower Alps, forests make up another large part of the potential (9 TWh). Since the forest line marks the approximate separation between lower and upper Alps, they have only a small contribution to the potential in the upper Alps. The Plateau follows the Alps with around 10 TWh, of which around 4 TWh are in the *other* zone, while the Jura may allow for the exploitation of almost 6 TWh of wind energy. Wetlands, which are strictly protected at the federal level and at the same time constitute only a small area outside of lakes and rivers, are neglected here. Across all parts of Switzerland, the *other* zone makes up 45% of the potential, followed by *forests* (30%) and *restricted* zones (25%).

## 5 Discussion

### 5.1 Methodological contributions

In this paper, we propose an adaptation of the spatio-temporal framework originally proposed in Amato et al. (2020b), adopting ELM ensembles to individually predict each spatial coefficient map resulting from the EOF decomposition of the spatio-temporal data. The variance estimates developed in Guignard et al. (2021) were used to extend the uncertainty quantification to the spatio-temporal framework. The prediction variance was estimated through a second model based on squared residuals after their log-transformation. The ELM based variance estimate of this second model was further used to back-transform the results. These developments were applied on hourly wind speed data for Switzerland. As shown in detail in "Appendix B in Electronic suuplementary material", the use of the regularised version of the ELM provides the opportunity to extract insightful information about the spatio-temporal model to understand its behaviour, but also to improve the explainability of the models in terms of data interpretability. In this specific case, those insights were also confirmed by the residual analysis.

The potential wind power generation was then estimated based on the modelled wind speed to assess renewable energy potential in Switzerland. As expected, the high variance propagated in the transition phase of the logistic function can lead to very uncertain predictions. An alternative way to estimate wind power may be to spatio-temporally model directly the transformed power data. However, a significant advantage of modelling the wind speed as a first step is that the obtained results do not depend on the choice of a specific turbine height and logistic parameters describing technical specificities of the turbine through the power curve. Hence, the power estimation can easily be updated to adapt to different choices of these parameters, generating multiple turbine scenarios to support decisions related to the turbine selection.

### 5.2 Practical contributions

The work presented here may contribute to the development of wind power in Switzerland in several ways. First, the hourly wind profiles, estimated for 10 years at a scale of $250 \times 250$ m$^2$ for the entire country, provide an exhaustive database for the modelling of potential future wind turbines in the Swiss electricity grid. The hourly temporal resolution hereby allows to assess the complementarity of wind power with other renewable resources such as solar photovoltaics (Dujardin et al. 2017; Zappa and van den Broek 2018), and to quantify the potential impact of an increased share of wind power on the stability of the electricity grid (Gupta et al. 2021; Bartlett et al. 2018).

Second, the analysis of the annual wind power generation potential (Sect. 4.3) may be set into context with the goals of the "Swiss Energy Perspectives", aiming at a wind power generation of 4.3 TWh by 2050 (BFE 2020). This target corresponds to an increase of the current production by a factor of 30 (S.F.I. for Energy 2018). With an average annual wind power potential of 4.4 GWh, this target may be achieved through the installation of around 1000 wind turbines. This target is rather low compared to other European countries (WindEurope Business intelligence 2021), potentially due to the large part of the country being covered by mountains, as well as strong societal and political concerns. The target of 4.3 TWh hence lies well within the potentials identified in Sect. 4.3, and may be achieved by realising less than 20% of the potential in the *other* zone.

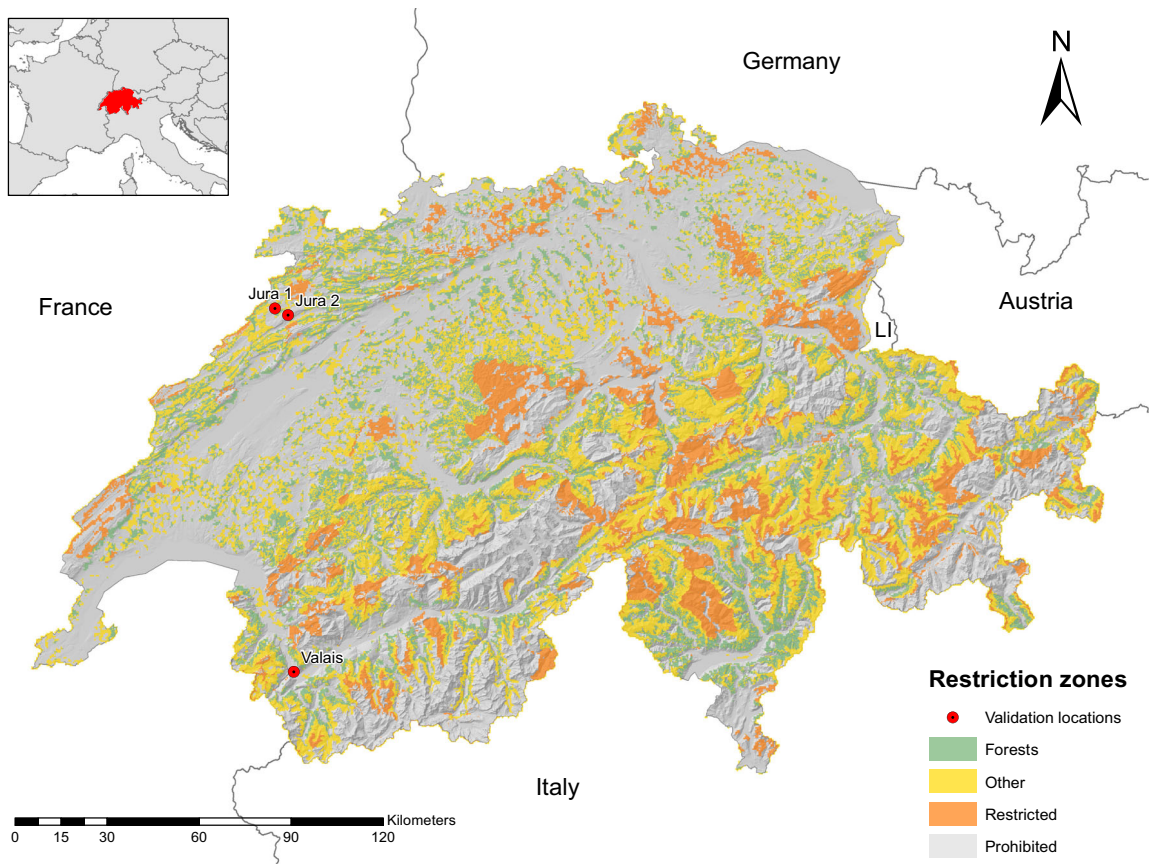

**Fig. 7** Spatial distribution of restriction zones. Restriction zones (see Table 3) within the case study region of Switzerland, and location of the validation sites (see Table 6)

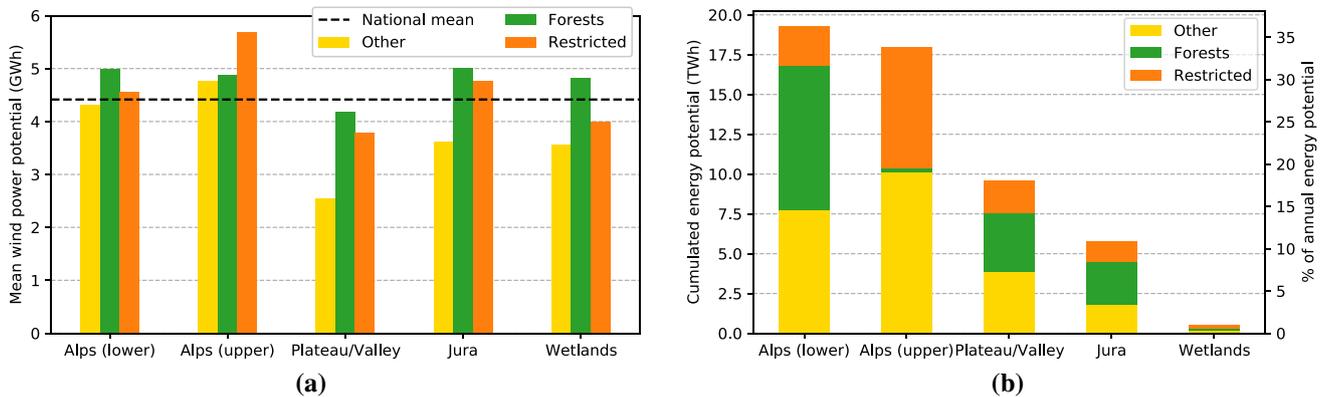

**Fig. 8** Annual potential wind power generation potential for Switzerland. Annual mean power generation per turbine (**a**) and annual total generation of all (virtual) turbines (**b**), for each part of Switzerland (bars) and for each restriction zone (colors)

Third, overlaying the information on wind power generation potential, the variance of this potential and the available area for turbine installation may serve to identify suitable areas for future wind farms in Switzerland. The variance plays a key role in this process, as potential wind

**Table 5** Technical characteristics of existing wind power installations of ∼ 100 m hub height (cf. Hertach and Schlegel 2020)

| Installation | Facility ID | Number of turbines | Construction year | Model | Hub height (m) | Diameter (m) | Rated speed (m/s) | Rated power (kW) |
| --- | --- | --- | --- | --- | --- | --- | --- | --- |
| Valais | COL | 1 | 2005 | Enercon E-70 | 100 | 71 | | 2000 |
| | MTG | 1 | 2008 | Enercon E-82 | 99 | 82 | 12 | 2000 |
| | CHA | 1 | 2012 | Enercon E-101 | 99 | 101 | 13 | 3000 |
| Jura 1 | PEU | 3 | 2010 | Enercon E-82 | 108 | 80 | 12 | 2300 |
| Jura 2 | MTC | 12 | 2010–2013 | Vestas V90 | 95 | 90 | 12 | 2000 |
| | MTC | 4 | 2016 | Vestas V112 | 94 | 112 | 12 | 3300 |

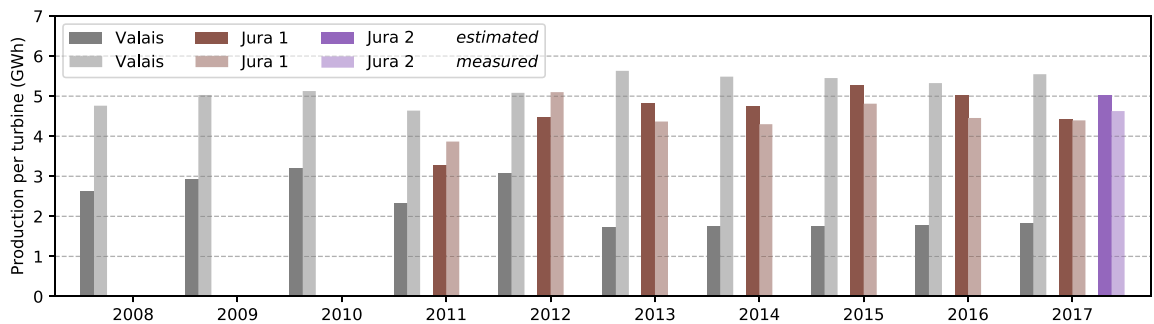

**Fig. 9** Validation of potential wind power generation. Estimated (non-transparent bars) and measured (semi-transparent bars) annual potential wind power generation for three existing installations (Valais, Jura 1, Jura 2)

farms in areas with low variance may allow for a higher planning reliability.

The work presented in this study is an assessment of the potential wind speeds and wind power generation. It does hence not represent an installation recommendation for wind turbines in a specific location, nor does it replace any local measurements in future wind projects. Instead, it is aimed to be used in studies of future electricity grids, by the scientific community or by energy planners, and to provide further insights for policy makers in the development of national renewable energy targets, while accounting for the need of protecting natural systems, often endangered by power plant expansions.

### 5.3 Validation and comparison to existing studies

To validate the proposed method, the estimated annual potential wind power generation is compared to measured electricity production from three wind power plants in Switzerland (see Fig. 7), two of which are located in the Jura, and one in the Rhone Valley (Valais). These are the only three of Switzerland's 40 wind power facilities with turbine heights around 100 m (90–110 m considered) with measured electricity generation before 2018. Table 5 provides an overview of the technical features of these power plants. As the installation "Jura 2" also contained several wind turbines of lower hub heights which were decommissioned between 2013 and 2016, for this installation only the data for 2017 can be used for the validation.

As Fig. 9 shows, the estimated annual production per turbine lies within ± 15% of the measured values for the two installations in the Jura (see also Table 6). For the turbines installed in the Rhone Valley, an underestimation of up to 69% is observed, particularly for the years after 2013. A part of this underestimation may be due to uncertainties in the roughness length, since these turbines are located at the boundaries of industrial areas. Furthermore, jet-like flows through the Rhone Valley, peaking at 200 m above ground in the warm summer months (Schmid et al. 2020), may lead to an increased wind speed at the modelled height of 100 m, which are not accounted for by the applied log-law. In the "Rhone knee", the corner of the Rhone Valley, these effects are particularly pronounced, which creates major difficulties for modelling the wind speeds in this particular region (Koller and Humar 2016).

Table 6 Percentage difference between the estimated annual potential wind power generation and the measured data (see Fig. 9) for three existing wind installations in Switzerland (Valais, Jura 1, Jura 2)

| Installation | 2008 | 2009 | 2010 | 2011 | 2012 | 2013 | 2014 | 2015 | 2016 | 2017 | Mean |
|---|---|---|---|---|---|---|---|---|---|---|---|
| Valais | −45 | −42 | −37 | −50 | −39 | −69 | −68 | −68 | −66 | −67 | −56 |
| Jura 1 | | | | −15 | −12 | 10 | 11 | 10 | 13 | 1 | 2 |
| Jura 2 | | | | | | | | | | 8 | 8 |

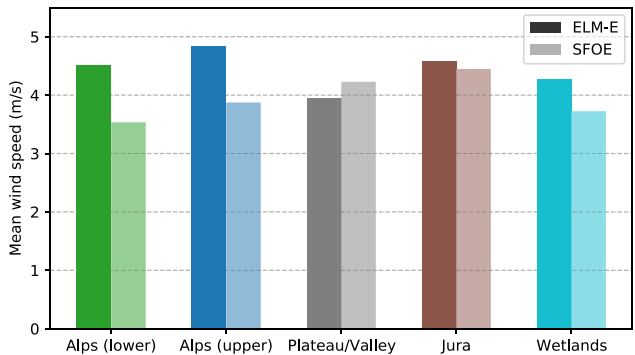

Fig. 10 Comparison to Swiss wind atlas. Annual mean wind speed for each part of Switzerland as estimated here (ELM-E, solid bars) and as estimated by the Swiss Federal Office of Energy (Koller and Humar 2016) (SFOE semi-transparent bars)

Finally, the rated power of the modelled wind turbine (3050 kWh) varies from the rated power of the turbines (see Table 5). As the rated power of the installations lies on average below that of the assumed wind turbine, the estimated power is expected to be above that of the measured data. However, this effect may be offset by different wind power curves that increase the generation at lower wind speeds. Due to the small size of the validation sample, these results cannot be considered to be representative.

In addition to the validation against measurement data, we compare the results to another existing estimation of annual wind speeds at 100 m height for Switzerland, published as part of the wind atlas of the Swiss Federal Office of Energy (SFOE) (Koller and Humar 2016). The wind atlas uses a computational fluid dynamics (CFD)-model based on the software WindSim, which computes annual average wind speeds at different heights and does not account for any temporal correlations or patterns, due to the high computational requirements of the model. As Fig. 10 shows, this approach estimates higher wind speeds than those estimated by SFOE, particularly in the alpine terrain. In the Jura and in the Plateau the difference between both estimates is small, whereby the estimated wind speeds in the Plateau is slightly lower in this study than estimated by SFOE.

The high complexity of the wind speed patterns in mountain terrain may be regarded as the primary reason for these differences, whereby the presented ELM-approach leads to higher estimates than the model based on CFD used by the SFOE. In addition to the computational methods, one of the main differences between these two estimations lies in the temporal resolution of the results. While the SFOE-estimate is based on average wind conditions (Koller and Humar 2016), this work yields results in hourly resolution. These hourly data may be used for example in studies of hybrid energy systems with high shares of wind power.

### 5.4 Limitations and further work

The estimation of the potential generation of wind turbines of 100 m hub height is limited by the data availability of wind speeds at 10 m only. This requires the use of physical and empirical formulas to estimate wind power generation, namely the log-law and the wind power curve. Propagating the variances through these formulas increases the variance of the estimated potential. The log-law further requires the estimation of roughness length, which is approximated from land use data, leading to further uncertainties. Additionally, wind phenomena occurring at the target height of 100 m, such as thermally induced winds in mountain valleys, are not taken into account through the extrapolation via the log-law (see Sect. 5.3), and can only be considered if wind measurements are available at 100 m height.

Future work may aim at a further validation and calibration of the proposed model by collecting and integrating hourly monitored data of wind speed and wind power generation at heights above 10 m, which are currently unavailable for Switzerland. The estimated generation, variance and available area may further be combined to develop a suitability indicator for wind power, accounting for these three factors. The hourly temporal resolution of the results allows to derive further indicators related to the intermittency of wind power. Finally, the proposed model may be expanded, at national scale or for particular areas of interest, to account for different hub heights and wind turbines. This is the main advantage of using the physical and empirical formulas mentioned above. Such a tool may be used to choose suitable turbine models to maximise the wind power output at a specific location.

# 6 Conclusions

In this paper we propose an estimation of hourly wind energy potential at the Swiss national scale. The application was developed using a newly-introduced framework enabling spatio-temporal prediction of data measured on irregularly spaced monitoring networks. A particular attention was paid to uncertainty quantification and its propagation throughout the entire modelling procedure. Particularly, 10 years of wind speed measurement collected at an hourly frequency on three sets of up to 208 monitoring stations. The data were interpolated using advanced spatio-temporal techniques, in order to estimate wind speed at unsampled locations. Then, the resulting wind field was used to estimate hourly wind power potential on a national scale on a reguar grid having a spatial resolution of 250 m.

The results showed that the wind power potential is highest in the mountain areas of the Alps and the Jura, of which the wind speeds in the Jura mountains have an overall lower variance. The conversion of wind speed to wind power through the power curve leads to high uncertainties whenever the wind speed is in the transition region of the logistically approximated power curve. Across Switzerland, we estimate an annual average power generation for turbines at 100 m hub height of 4.4 GWh, with intra-annual variations by up to 15–20%. A validation has shown that the estimated potential deviates by less than 15% from the measured annual electricity yield in the Jura, while there are some limitations for the estimation of wind power in the Rhone valley.

The virtual installation of wind turbines on all available area with a spacing of 1.6 km$^2$ yields a potential 12,000 turbines on around half of the Swiss terrain. About 1000 of these turbines would be sufficient to fulfil the targets of the Swiss energy perspectives of 4.3 TWh by 2050, which may be realised by installing wind turbines exclusively in areas without identified restrictions.

The high spatio-temporal resolution of the results, as hourly values for 10 years for pixels of $250 \times 250$ m$^2$, allows to integrate the results in increasingly complex national energy systems models aiming at the optimization of renewable energy use across Switzerland. A combination of the wind power potential, its uncertainty and the available area for turbine installation further enables the assessment of the suitability of different areas for future wind projects. Further methodological development may lead to the definition of ELM confidence and prediction intervals for the estimated wind power. The current work aims to support the development of wind power as part of a fully renewable future energy system in Switzerland.

**Supplementary Information** The online version contains supplementary material available at https://doi.org/10.1007/s00477-022-02219-w.

**Acknowledgements** The research presented in this paper was supported by the National Research Program 75 "Big Data" (PNR75, Project No. 167285 "HyEnergy") of the Swiss National Science Foundation (SNSF).

**Author contributions** FA and FG conceived the main conceptual ideas. FG preprocessed the data, and developed the theoretical formalism FA designed the experiments and performed the calculations. AW postprocessed, analyzed and validated the computational results. FA, FG and AW wrote the original draft and discussed the results. NM, JLS, MK, carried out the supervision and funding acquisition. All authors reviewed the manuscript and gave final approval for its publication.

**Funding** Open access funding provided by EPFL Lausanne. Funding was provied by Schweizerischer Nationalfonds zur Förderung der Wissenschaftlichen Forschung (Grant No. 167285).

**Data availability** The results produced in this paper with respect to the estimation of the average yearly wind speed and of the wind power potential for Switzerland, at a spatial resolution of $250 \times 250$ m and over the period from 2008 to 2017 are available at Amato et al. (2021).

## Compliance with ethical standards

Stochastic Environmental Research and Risk Assessment

Stochastic Environmental Research and Risk Assessment

**Publisher's Note** Springer Nature remains neutral with regard to jurisdictional claims in published maps and institutional affiliations.

## Authors and Affiliations

Federico Amato[1]  · Fabian Guignard[2] · Alina Walch[3] · Nahid Mohajeri[4] · Jean-Louis Scartezzini[3] · Mikhail Kanevski[5]

✉ Federico Amato
federico.amato@epfl.ch

Fabian Guignard
fabian.guignard@stat.unibe.ch

Alina Walch
alina.walch@epfl.ch

Nahid Mohajeri
nahid.mohajeri.09@ucl.ac.uk

Jean-Louis Scartezzini
jean-louis.scartezzini@epfl.ch

Mikhail Kanevski
mikhail.kanevski@unil.ch

[1] Swiss Data Science Centre, École polytechnique fédérale de Lausanne (EPFL) and Eidgenössische Technische Hochschule Zurich (ETH), Zurich, Switzerland

[2] Institute of Mathematical Statistics and Actuarial Science, University of Bern, Bern, Switzerland

[3] Solar Energy and Building Physics Laboratory, Ecole Polytechnique Fédérale de Lausanne, Lausanne, Switzerland

[4] Institute of Environmental Design and Engineering, Bartlett School of Environment, Energy and Resources, University College London, London, UK

[5] Institute of Earth Surface Dynamics, University of Lausanne, Lausanne, Switzerland